%% file: main.tex
\documentclass{article}

\usepackage[dblblindworkshop, final]{neurips_2026}
\workshoptitle{Resource-Aware Agentic AI}

\usepackage[utf8]{inputenc} 
\usepackage[T1]{fontenc}    
\usepackage{hyperref}       
\hypersetup{hidelinks}
\usepackage{url}            
\usepackage{booktabs}       
\usepackage{amsfonts}       
\usepackage{nicefrac}       
\usepackage{microtype}      
\usepackage[table]{xcolor}  
\usepackage{amsmath}
\usepackage{makecell}

\usepackage{multirow}       
\usepackage{graphicx}       
\usepackage{float}          

\providecommand{\hmethod}{HyLaMAS}
\providecommand{\selectkv}{SelectKV}

\title{Selective and Hybrid-State Communication for Latent Multi-Agent Systems}

\author{%
  Miles Liao\thanks{Equal contribution.} \\
  Hopkins School \\
  \texttt{mileshliao@gmail.com}
  \And
  Aaron Smieszek\footnotemark[1] \\
  Independent Researcher \\
  \texttt{aaron.smieszek@gmail.com}
  \And
  Yuhao Ge \\
  Independent Researcher \\
  \texttt{geyuhao33@gmail.com}
}

\begin{document}

\maketitle
\raggedbottom

\begin{abstract}
Efficient communication remains a major bottleneck in large language model
multi-agent systems. Latent communication reduces repeated natural-language
exchange by transferring internal model states between agents, but these
transfers remain costly and are typically designed around conventional KV-cache
representations. We address this problem from two complementary perspectives.
We introduce \textbf{SelectKV}, a boundary-aware method that selectively
communicates KV states while protecting inherited context, and
\textbf{\hmethod}, which extends latent collaboration to hybrid linear-attention
models by communicating both KV caches and recurrent states through a shared
workspace. Across six benchmarks, SelectKV reduces communicated KV positions
and logical KV payload by 9.7--9.9\%, with accuracy unchanged on two benchmarks,
within one percentage point on three, and six points lower on HumanEval+.
Across Qwen3.5 model scales, \hmethod\ reduces decoded output tokens by
53--80\% and achieves 1.16--2.78$\times$ speedups relative to text-based
multi-agent communication. Together, these independently evaluated methods represent two
complementary approaches to efficient latent-state communication
across conventional Transformer and hybrid model architectures.
\end{abstract}

\section{Introduction}

Large language model (LLM) multi-agent systems improve complex reasoning by
allowing multiple agents to iteratively exchange and refine information
\citep{du2023improvingfactualityreasoninglanguage, liang2024encouragingdivergentthinkinglarge}. Most systems communicate through
natural language, requiring each agent to decode its internal state into text
and the next agent to tokenize and process that text again \citep{zhang2024chainagentslargelanguage, wu2023autogenenablingnextgenllm, du2023improvingfactualityreasoninglanguage}. Latent
communication instead transfers internal representations directly, reducing
this decode--re-encode overhead \citep{pham2024letmodelsspeakciphers, fu2026cachetocachedirectsemanticcommunication, ramesh2025communicatingactivationslanguagemodel,zou2026latentcollaborationmultiagentsystems}.

However, latent communication introduces a new systems question: \emph{what
internal state should be transmitted between agents?} Existing approaches
largely rely on transferring KV-cache state
\citep{zou2026latentcollaborationmultiagentsystems,shi2026kvcommenablingefficientllm,ye2025kvcommonlinecrosscontextkvcache},
which can contain redundant information as trajectories grow. Prior KV-cache
compression methods show that only a subset of cached states may be important
for future decoding \citep{liu2023scissorhandsexploitingpersistenceimportance,
zhang2023h2oheavyhitteroracleefficient,xiao2024efficientstreaminglanguagemodels},
but these methods primarily target single-model inference rather than
inter-agent state transfer. At the same time, emerging hybrid
linear-attention architectures such as Qwen3.5 combine full-attention KV
caches with complete Gated DeltaNet (GDN) inference states
\citep{yang2025gateddeltanetworksimproving,qwen35blog}, so a KV-only handoff is
no longer sufficient.

We study two complementary, independently evaluated approaches to latent-state communication. \textbf{SelectKV} selects newly communicable KV states while preserving inherited context. \textbf{\hmethod{}} extends LatentMAS to hybrid architectures by transferring full-attention KV caches together with the recurrent and causal-convolution states of Gated DeltaNet layers.

Together, SelectKV and \hmethod\ study efficient latent-state handoff from two
complementary perspectives: selecting the information that should be
communicated and supporting the heterogeneous states required by modern model
architectures. Across reasoning, knowledge, and code-generation benchmarks,
our results show that latent communication can be made more selective while
largely preserving downstream performance, and that hybrid-state transfer can
substantially reduce the overhead of text-based multi-agent communication.

\textbf{Key Contributions:}
\begin{itemize}
    \item \textbf{Selective inter-agent KV communication.}
    We introduce SelectKV, a boundary-aware KV selection mechanism for latent
    multi-agent handoff that preserves inherited context while selecting
    newly generated KV states. Ablations examine boundary protection and
    matched-budget selection baselines.

    \item \textbf{Hybrid-state latent communication.}
    We introduce \hmethod, a shared latent workspace for hybrid
    linear-attention models that transfers both full-attention KV caches and
    recurrent linear-attention states between agents.

    \item \textbf{Multi-benchmark evaluation.}
    Across six reasoning, knowledge, and code-generation benchmarks, SelectKV
    reduces KV communication while largely preserving full-cache LatentMAS
    performance, while \hmethod\ reduces generated-token and wall-clock
    overhead relative to text-based multi-agent communication.
\end{itemize}

\section{Background and Related Work}

\subsection{Multi-Agent Systems and Latent Communication}

LLM-based multi-agent systems commonly coordinate through natural-language
messages, requiring intermediate reasoning to be decoded into text and
subsequently re-encoded and processed by other agents
\citep{zhang2024chainagentslargelanguage,
wu2023autogenenablingnextgenllm,
du2023improvingfactualityreasoninglanguage}.
Latent communication instead enables agents to exchange internal
representations directly. In particular, LatentMAS transfers latent
representations and KV-cache state across agents, allowing a receiving agent
to continue from a preceding reasoning trajectory without relying exclusively
on textual messages
\citep{zou2026latentcollaborationmultiagentsystems}.

For a Transformer with $L$ layers, the KV cache can be written as
\begin{equation}
\mathbf{C} =
\{(\mathbf{K}^{(l)},\mathbf{V}^{(l)})\}_{l=1}^{L},
\end{equation}
where $\mathbf{K}^{(l)}$ and $\mathbf{V}^{(l)}$ denote the cached key and value
states at layer $l$. Because the cache grows linearly with sequence length,
transferring the full accumulated state between agents can become increasingly
costly as multi-agent reasoning trajectories grow. Recent work has therefore
explored selective, compressed, or reusable KV communication across agents
\citep{shi2026kvcommenablingefficientllm,
ye2025kvcommonlinecrosscontextkvcache,
li2026latentleadsbetterrelay}.

\subsection{Redundancy in KV Caches}

Prior work on efficient Transformer inference has shown that cached positions
contribute unequally to subsequent computation. H$_2$O identifies attention
``heavy hitters'' and preferentially retains frequently attended positions
\citep{zhang2023h2oheavyhitteroracleefficient}; Scissorhands exploits the
persistence of token importance across generation
\citep{liu2023scissorhandsexploitingpersistenceimportance}; and StreamingLLM
demonstrates the importance of preserving initial attention-sink tokens during
long-context inference
\citep{xiao2024efficientstreaminglanguagemodels}. Together, these findings
suggest substantial redundancy in accumulated KV state and motivate selective
retention rather than uniform preservation of all cached positions.

Related work extends this principle from single-model inference to inter-agent
communication. KVComm studies selective KV sharing between LLM agents
\citep{shi2026kvcommenablingefficientllm}, while cross-context KV-cache
communication enables cached context to be reused across agents
\citep{ye2025kvcommonlinecrosscontextkvcache}. Orthogonal BackFill (OBF)
compresses latent multi-agent communication while reconstructing information
from discarded states
\citep{li2026latentleadsbetterrelay}. $k$NN-latentMAS ranks cached states using
semantic similarity to downstream context and retains the most relevant KV
entries \citep{knnlatentmas}. These approaches establish KV selection and
compression as mechanisms for reducing latent communication while preserving
task-relevant state.

\subsection{Emerging Architectures Beyond Standard Attention}

Latent communication methods have primarily been developed for standard
attention-based Transformers, where an agent's accumulated reasoning state can
be represented through a conventional KV cache
\citep{zou2026latentcollaborationmultiagentsystems,
fu2026cachetocachedirectsemanticcommunication,
shi2026kvcommenablingefficientllm,
ye2025kvcommonlinecrosscontextkvcache}.
Recent language-model architectures, however, increasingly incorporate
recurrent, state-space, or linear-attention mechanisms whose inference state
differs from the KV cache of softmax attention
\citep{gu2024mambalineartimesequencemodeling,
dao2024transformersssmsgeneralizedmodels,
yang2024gatedlinearattentiontransformers,
yang2025gateddeltanetworksimproving}.

Some architectures combine recurrent or linear-attention layers with
conventional attention layers
\citep{lieber2024jambahybridtransformermambalanguage,qwen35blog},
while others replace standard attention more extensively
\citep{gu2024mambalineartimesequencemodeling,
zuo2024falconmambacompetitiveattentionfree}.
Their inference state may therefore include recurrent states in addition to,
or instead of, conventional key--value caches. As these architectures become
more prevalent, latent multi-agent communication must account for internal
state representations that cannot be described solely by Transformer KV
caches.

\section{Our Proposed Method}

We formulate efficient latent collaboration as the problem of determining
\emph{what internal state should cross an agent boundary}. We address two
complementary aspects of this problem. \textbf{SelectKV} reduces
communication in Transformer-based latent multi-agent systems by selectively
retaining newly generated KV states at each agent boundary. \textbf{\hmethod}
extends latent collaboration to hybrid architectures whose inference state
contains both conventional attention KV caches and recurrent
linear-attention states.

\subsection{SelectKV: Boundary-Aware KV Selection}

Consider the Transformer KV cache $\mathbf{C}$ defined in Section~2.1 at
an inter-agent handoff.

Suppose the cache contains $T$ positions at the current handoff, of which the
first $T_{\mathrm{prev}}$ were inherited from preceding agents. We partition
the cache into
\begin{equation}
\mathcal{I}_{\mathrm{prev}}
=
\{1,\ldots,T_{\mathrm{prev}}\},
\qquad
\mathcal{I}_{\mathrm{new}}
=
\{T_{\mathrm{prev}}+1,\ldots,T\}.
\label{eq:selectkv-boundary}
\end{equation}
SelectKV protects $\mathcal{I}_{\mathrm{prev}}$ from repeated compression
and selects positions from $\mathcal{I}_{\mathrm{new}}$ using relevance
and local KV novelty.

\subsubsection{Relevance Scoring}

Motivated by similarity-based KV retrieval \citep{knnlatentmas},
let $\bar{\mathbf{k}}_i$ be the layer- and head-averaged key at position
$i$, and $\bar{\mathbf{q}}$ a sender-derived relevance query in the same
space. We score candidate positions by cosine similarity:
\begin{equation}
s_i^{\mathrm{rel}}
=
\cos
\left(
\bar{\mathbf{k}}_i,
\bar{\mathbf{q}}
\right)
=
\frac{
\bar{\mathbf{k}}_i^{\top}\bar{\mathbf{q}}
}{
\left\|\bar{\mathbf{k}}_i\right\|_2
\left\|\bar{\mathbf{q}}\right\|_2
}.
\label{eq:selectkv-relevance}
\end{equation}

The query $\bar{\mathbf{q}}$ is the sending agent's final hidden state
after latent rollout, reduced to the key-head dimension by averaging
equal-sized chunks (or truncating if needed). Keys are averaged across
heads and layers; both representations are $\ell_2$-normalized.
The query is available before handoff, without a receiving-agent
forward pass. Higher scores indicate greater alignment with this
sender-derived proxy for downstream relevance.

\subsubsection{KV-Novelty Scoring}

To favor distinct information, we score local KV novelty across the
full cache of $T$ positions:
\begin{equation}
s_i^{\mathrm{nov}} =
\begin{cases}
1-\cos(\bar{\mathbf{k}}_1,\bar{\mathbf{k}}_2), & i=1,\\[2pt]
1-\dfrac{\cos(\bar{\mathbf{k}}_i,\bar{\mathbf{k}}_{i-1})
+\cos(\bar{\mathbf{k}}_i,\bar{\mathbf{k}}_{i+1})}{2},
& 1<i<T,\\[6pt]
1-\cos(\bar{\mathbf{k}}_T,\bar{\mathbf{k}}_{T-1}), & i=T.
\end{cases}
\label{eq:selectkv-novelty}
\end{equation}
Only positions in $\mathcal{I}_{\mathrm{new}}$ are eligible for selection.
Interior scores compare both neighbors, including the inherited boundary
neighbor when $i=T_{\mathrm{prev}}+1$; endpoints use one neighbor.
Higher scores indicate less redundancy. Unlike attention-based methods
\citep{zhang2023h2oheavyhitteroracleefficient,
liu2023scissorhandsexploitingpersistenceimportance},
this score requires no explicit attention extraction.

\subsubsection{Boundary-Aware Selection}

Let $N=|\mathcal{I}_{\mathrm{new}}|$ and let $r\in(0,1]$ be the
target retention ratio. The new-region selection budget is
\begin{equation}
n=\min\bigl(N,\lceil rN\rceil\bigr).
\label{eq:selectkv-budget}
\end{equation}

SelectKV prioritizes the overlap of the descending relevance and
novelty rankings, then alternates between their remaining top positions.
A small set of recent positions is protected within the budget
\citep{xiao2024efficientstreaminglanguagemodels}.
For the selected set $\mathcal{S}_r\subseteq\mathcal{I}_{\mathrm{new}}$,
$|\mathcal{S}_r|\leq n$, the transmitted positions are
\begin{equation}
\mathcal{I}_{\mathrm{send}}
=\mathcal{I}_{\mathrm{prev}}\cup\mathcal{S}_r.
\label{eq:selectkv-send}
\end{equation}
Both keys and values are retained at every layer for these positions.
We use $r=0.90$, chosen from preliminary retention experiments.

\subsection{\hmethod{}: Hybrid-State Latent Communication}

Whereas SelectKV reduces the amount of KV state communicated between
agents, HyLaMAS addresses a different challenge: enabling latent
communication when the model's inference state is not represented
solely by a KV cache.
Existing latent multi-agent communication methods such as LatentMAS primarily
operate over Transformer latent states and KV caches
\citep{zou2026latentcollaborationmultiagentsystems}. Hybrid architectures,
however, may maintain fundamentally different inference states across
different layer types. Qwen3.5, for example, interleaves Gated DeltaNet (GDN)
linear-attention layers with conventional full-attention layers
\citep{qwen35blog}. A KV-only handoff therefore cannot preserve the complete
inference state of such a model.

We introduce \textbf{\hmethod}, which communicates the layer-native inference
states across agent boundaries through a shared in-memory cache.
Rather than projecting one KV cache into another
\citep{fu2026cachetocachedirectsemanticcommunication}, \hmethod{} preserves
the appropriate state representation for each layer type.

Let $\mathcal{A}$ and $\mathcal{G}$ denote the sets of full-attention and GDN
layer indices, respectively. At agent boundary $i$, we define the shared
hybrid inference state as
\begin{equation}
\mathcal{H}_i
=
\left\{
\left(
\mathbf{K}^{(\ell)}_i,
\mathbf{V}^{(\ell)}_i
\right)
\right\}_{\ell\in\mathcal{A}}
\cup
\left\{
\left(
\mathbf{C}^{(\ell)}_i,
\mathbf{S}^{(\ell)}_i
\right)
\right\}_{\ell\in\mathcal{G}},
\label{eq:hybrid-state}
\end{equation}
where $i$ indexes an agent boundary and $\ell$ indexes a model layer.

For a full-attention layer $\ell\in\mathcal{A}$,
$\mathbf{K}^{(\ell)}_i$ and $\mathbf{V}^{(\ell)}_i$ contain the cached key and
value projections accumulated before boundary $i$. For a GDN layer
$\ell\in\mathcal{G}$, $\mathbf{S}^{(\ell)}_i$ denotes the recurrent matrix
state summarizing preceding computation, while
$\mathbf{C}^{(\ell)}_i$ contains the recent projected inputs required by the
layer's causal convolution
\citep{yang2025gateddeltanetworksimproving}.

The state components have shapes
\begin{align}
\mathbf{K}^{(\ell)}_i,\mathbf{V}^{(\ell)}_i
&\in
\mathbb{R}^{
B\times H_{\mathrm{kv}}\times T_i\times d_{\mathrm{attn}}
},
&& \ell\in\mathcal{A},
\\
\mathbf{S}^{(\ell)}_i
&\in
\mathbb{R}^{
B\times H_v\times d_k\times d_v
},
&& \ell\in\mathcal{G},
\\
\mathbf{C}^{(\ell)}_i
&\in
\mathbb{R}^{
B\times
(2d_k^{\mathrm{tot}}+d_v^{\mathrm{tot}})
\times\kappa
},
&& \ell\in\mathcal{G},
\end{align}
where $B$ is the batch size, $T_i$ is the number of cached positions at
boundary $i$, $H_{\mathrm{kv}}$ is the number of key/value heads in a
full-attention layer, and $d_{\mathrm{attn}}$ is the corresponding head
dimension. For GDN layers, $H_v$ is the number of value heads, $d_k$ and
$d_v$ are the key and value head dimensions,
$d_k^{\mathrm{tot}}=H_kd_k$,
$d_v^{\mathrm{tot}}=H_vd_v$, and $\kappa$ denotes the causal-convolution
kernel width.

Given the shared state $\mathcal{H}_i$, the next agent continues computation
from that state and produces an updated state
\begin{equation}
\mathcal{H}_{i+1}
=
F_{i+1}
\left(
\mathcal{H}_i,
\mathbf{x}_{i+1}
\right),
\label{eq:hybrid-transition}
\end{equation}
where $\mathbf{x}_{i+1}$ denotes the next agent's prompt and
$F_{i+1}$ denotes its model computation. Consequently, both recurrent
linear-attention state and conventional attention KV state persist across
agent boundaries rather than being reconstructed from decoded text.

Following the LatentMAS interaction structure, each non-Judger agent appends
its role-conditioned prompt and performs $m$ latent rollout steps without
decoding an intermediate textual message. The next agent continues from the shared hybrid state. The Judger receives the final accumulated state,
appends its prompt, and directly decodes the task response. Padding-aware
position and attention masks maintain independence among examples when
batched.

This formulation generalizes latent collaboration from Transformer-only KV
communication to architectures with heterogeneous inference states. In
particular, the recurrent GDN component has fixed size with respect to
sequence length, whereas the full-attention KV component continues to grow
with the number of cached positions. We analyze the communication implications
of these two state channels in Appendix~\ref{sec:ablations}.
\input{tables/select_kv_smaller_results}
\section{Evaluations}
\subsection{Experimental Setup}
Following the LatentMAS evaluation method \citep{zou2026latentcollaborationmultiagentsystems}, our two methods, \textbf{SelectKV} and \textbf{\hmethod}{}, were evaluated across six benchmarks that can be categorized into three groups: (i) Math and Science: GSM8K \citep{cobbe2021trainingverifierssolvemath} and MedQA \citep{jin2020diseasedoespatienthave}; (ii) Commonsense Reasoning: ARC-Easy \citep{clark2018thinksolvedquestionanswering} and ARC-Challenge \citep{clark2018thinksolvedquestionanswering}; (iii) Code Generation:  MBPP-Plus \citep{liu2023codegeneratedchatgptreally} and HumanEval-Plus \citep{liu2023codegeneratedchatgptreally}.

We evaluate 100 examples per benchmark using greedy decoding. For accuracy, we report 95\% percentile bootstrap confidence intervals
computed using 10,000 resamples over the $n=100$ evaluation examples. The latent protocols follow a sequential Planner--Critic--Refiner--Judger chain, with only the Judger decoding text. Model-specific rollout budgets and generation caps are specified below. Hardware, software, realignment, and measurement details are provided in Appendix~\ref{app:setup}.

\subsection{SelectKV Compression}

We compare boundary-aware \selectkv{}-90 (90\% retention of newly
communicable KV positions) against uncompressed LatentMAS on Qwen3-4B
with 10 latent steps, greedy decoding, batch size 1, and seed 42.
Inference time is total experiment wall-clock time per example,
including agent inference, query processing, scoring, ranking,
and cache pruning, but not isolated inter-device transfer latency.

Table~\ref{tab:selectkv-results} shows 9.7--9.9\% reductions in
communicated KV positions and logical payload across all six benchmarks.
Accuracy is unchanged on ARC-Easy and ARC-Challenge, one point lower
on GSM8K, MedQA, and MBPP+, and six points lower on HumanEval+
(89\% to 83\%) despite a 9.8\% payload reduction. Thus, the fixed
retention policy has task-dependent accuracy costs. Inference is faster
on four benchmarks, with little change on GSM8K and HumanEval+.
At $n=100$, bootstrap intervals are wide and overlap across methods.

\subsection{\hmethod\ - Hybrid-State LatentMAS}
We evaluate \hmethod\ against the six benchmarks outlined above in order to determine whether hybrid-state latent collaboration reduces decoded tokens and wall-clock time relative to text-based multi-agent systems. We used Qwen3.5-2B, Qwen3.5-4B, and Qwen3.5-9B \citep{qwen35blog}. For the comparison, we evaluate \hmethod\ against: (i) Single LLM Agents, where one LLM strictly performs standard autoregressive generation with token-level decoding, and (ii) Sequential text-based MAS (TextMAS), following the chain-of-agents design \citep{zhang2024chainagentslargelanguage}.
\input{tables/main_results}
For each benchmark, 40 latent steps were generated per non-Judger agent during reasoning. Additionally, in order to ensure reproducibility and fairness, greedy decoding was used (Temperature = 0). For each benchmark, different token caps were used: For ARC-Easy, ARC-Challenge, and GSM8K, 2048 was the maximum token budget per agent allowed. For MedQA, MBPP+, and HumanEval+ it was 4096 tokens.

Across the six benchmarks, HyLaMAS generates 53--74\% fewer
decoded output tokens than TextMAS and achieves
$1.16$--$2.15\times$ faster inference. Most observed accuracy differences are small relative to sampling uncertainty at $n$ = 100; we therefore treat them as unresolved rather than evidence of statistical equivalence. Larger positive differences are observed on GSM8K at 4B and 9B and on HumanEval+ at 4B.

To determine whether \hmethod\ remains effective at smaller model scales, we repeated the evaluation using Qwen3.5-2B under the same four-agent, 40 latent-step protocol. As shown in Table~\ref{tab:smallmodel-results}, \hmethod\ reduces generated output tokens by 74.1-80.0\% and achieves a 2.10-2.78$\times$ speedup relative to TextMAS across all six benchmarks. Its accuracy is never lower than that of TextMAS, with differences ranging from 0 to 13 percentage points, with the largest gain on GSM8K (80\% vs. 67\%). These results extend \hmethod's primary efficiency finding to a 2B hybrid backbone: transferring the hybrid internal state avoids much decoding and re-encoding overhead associated with textual communication, without an observed reduction in task accuracy.

\hmethod{} reduces decoded output tokens and inference time relative to TextMAS across the evaluated settings, while accuracy changes vary by task and model scale. These benefits do not imply consistent improvements over single-agent inference. For example, on Qwen3.5-9B MBPP+, Single achieves 85\% accuracy and \hmethod{} achieves 76\%. The results therefore characterize the trade-offs of replacing textual multi-agent communication with hybrid-state communication under the evaluated protocols.

\input{tables/small_model_results}

\section{Conclusion}

We study complementary approaches to latent-state handoff.
\selectkv{} reduces communicated KV positions and payload by 9.7--9.9\%
across six benchmarks, with unchanged accuracy on two, one-point
decreases on three, and a six-point decrease on HumanEval+.
Matched-budget ablations favor informed selection over random and
recency-only baselines; boundary comparisons suggest a better
accuracy--compression trade-off when inherited context is preserved.
\hmethod\ extends latent handoff to hybrid architectures and reduces
generated tokens and inference time relative to text-based multi-agent
communication across evaluated Qwen3.5 scales.

\section{Limitations and Future Directions}

We evaluate only 100 examples per benchmark and one SelectKV model
scale (Qwen3-4B); broader benchmarks, model scales, and longer agent
chains are needed to assess generalization. SelectKV and \hmethod\
are evaluated independently, not as a combined selective hybrid-state
system. Logical payload reductions do not establish physical network
savings or proportional end-to-end speedups: selection incurs
computation, and hybrid-state handoff requires state management.
Future work should evaluate distributed transfer and adaptive
communication policies.



\bibliographystyle{plainnat}
\bibliography{sources.bib}
\clearpage

\appendix
\input{appendix_revised}

\end{document}

%% file: tables/select_kv_smaller_results.tex

\providecommand{\selectkv}{SelectKV}

\definecolor{winB}{HTML}{BFE3BF}

\providecommand{\up}[1]{$\uparrow$\,#1}
\providecommand{\dn}[1]{$\downarrow$\,#1}
\providecommand{\spd}[1]{$\times$#1}
\providecommand{\ours}[1]{\cellcolor{winB}\textbf{#1}}
\providecommand{\acc}[2]{\makecell{#1\\[-1pt]\scriptsize[#2]}}

\begin{table}[!b]
  \caption{
    Sequential multi-agent results on Qwen3-4B with 10 latent reasoning steps
    and $n=100$ examples per benchmark.
    LM denotes uncompressed LatentMAS; SKV-90 denotes boundary-aware
    \selectkv{} with a 90\% retention ratio applied to the newly communicable
    KV region while preserving inherited context.
    Metrics report accuracy (\%), generated output tokens, communicated KV
    positions, logical KV payload (MiB), and inference time (s), each averaged
    per example.
    Accuracy is reported with 95\% percentile bootstrap confidence intervals
    in brackets, computed using 10,000 resamples over evaluation examples.
    $\Delta$ reports the accuracy difference in percentage points, relative
    change for Token, KV Pos., and Payload, and the LM/SKV-90 inference-time
    ratio.
    Green shading indicates values favorable to SKV-90.
  }
  \label{tab:selectkv-results}

  \centering
  \small
  \setlength{\tabcolsep}{2.4pt}
  \renewcommand{\arraystretch}{1.08}

  \begin{tabular}{@{}l rrr rrr rrr@{}}
    \toprule
    & \multicolumn{3}{c}{\textbf{ARC-E}}
    & \multicolumn{3}{c}{\textbf{ARC-C}}
    & \multicolumn{3}{c}{\textbf{GSM8K}} \\
    \cmidrule(lr){2-4}
    \cmidrule(lr){5-7}
    \cmidrule(l){8-10}

    \textbf{Metric}
    & \textbf{LM} & \textbf{SKV-90} & \textbf{$\Delta$}
    & \textbf{LM} & \textbf{SKV-90} & \textbf{$\Delta$}
    & \textbf{LM} & \textbf{SKV-90} & \textbf{$\Delta$} \\
    \midrule

    Acc.
    & \textbf{\acc{99.0}{97.0, 100.0}}
    & \textbf{\acc{99.0}{97.0, 100.0}}
    & 0.0
    & \textbf{\acc{90.0}{84.0, 95.0}}
    & \textbf{\acc{90.0}{84.0, 95.0}}
    & 0.0
    & \textbf{\acc{94.0}{89.0, 98.0}}
    & \acc{93.0}{88.0, 98.0}
    & \dn{1.0} \\

    Token
    & 461.89 & \ours{429.23} & \ours{\dn{7.1\%}}
    & 561.03 & \ours{515.62} & \ours{\dn{8.1\%}}
    & \textbf{622.16} & 626.78 & \up{0.7\%} \\

    KV Pos.
    & 1023.04 & \ours{923.42} & \ours{\dn{9.7\%}}
    & 1099.60 & \ours{992.28} & \ours{\dn{9.8\%}}
    & 1068.46 & \ours{964.29} & \ours{\dn{9.7\%}} \\

    Payload
    & 143.87 & \ours{129.86} & \ours{\dn{9.7\%}}
    & 154.63 & \ours{139.54} & \ours{\dn{9.8\%}}
    & 150.25 & \ours{135.60} & \ours{\dn{9.7\%}} \\

    Time
    & 16.31 & \ours{14.83} & \ours{\spd{1.10}}
    & 19.33 & \ours{17.81} & \ours{\spd{1.09}}
    & \textbf{21.31} & 21.87 & \spd{0.97} \\

    \midrule

    & \multicolumn{3}{c}{\textbf{MedQA}}
    & \multicolumn{3}{c}{\textbf{MBPP+}}
    & \multicolumn{3}{c}{\textbf{HumanEval+}} \\
    \cmidrule(lr){2-4}
    \cmidrule(lr){5-7}
    \cmidrule(l){8-10}

    \textbf{Metric}
    & \textbf{LM} & \textbf{SKV-90} & \textbf{$\Delta$}
    & \textbf{LM} & \textbf{SKV-90} & \textbf{$\Delta$}
    & \textbf{LM} & \textbf{SKV-90} & \textbf{$\Delta$} \\
    \midrule

    Acc.
    & \textbf{\acc{63.0}{53.0, 72.0}}
    & \acc{62.0}{52.0, 71.0}
    & \dn{1.0}
    & \textbf{\acc{72.0}{63.0, 81.0}}
    & \acc{71.0}{62.0, 80.0}
    & \dn{1.0}
    & \textbf{\acc{89.0}{83.0, 95.0}}
    & \acc{83.0}{75.0, 90.0}
    & \dn{6.0} \\

    Token
    & 1212.50 & \ours{961.38} & \ours{\dn{20.7\%}}
    & 1527.42 & \ours{1436.22} & \ours{\dn{6.0\%}}
    & \textbf{1645.08} & 1761.82 & \up{7.1\%} \\

    KV Pos.
    & 2103.52 & \ours{1895.90} & \ours{\dn{9.9\%}}
    & 1545.22 & \ours{1393.25} & \ours{\dn{9.8\%}}
    & 1589.56 & \ours{1433.45} & \ours{\dn{9.8\%}} \\

    Payload
    & 295.81 & \ours{266.61} & \ours{\dn{9.9\%}}
    & 217.30 & \ours{195.93} & \ours{\dn{9.8\%}}
    & 223.53 & \ours{201.58} & \ours{\dn{9.8\%}} \\

    Time
    & 40.56 & \ours{31.36} & \ours{\spd{1.29}}
    & 51.33 & \ours{46.61} & \ours{\spd{1.10}}
    & \textbf{56.09} & 56.86 & \spd{0.99} \\

    \bottomrule
  \end{tabular}
\end{table}

%% file: tables/main_results.tex
%
\providecommand{\method}{HyLaMAS}
\definecolor{winB}{HTML}{BFE3BF}
\providecommand{\up}[1]{$\uparrow$\,#1}
\providecommand{\dn}[1]{$\downarrow$\,#1}
\providecommand{\spd}[1]{$\times$#1}
\providecommand{\gain}[1]{\cellcolor{winB}\textbf{#1}}

\begin{table}[!b]
  \caption{Sequential multi-agent results on Qwen3.5-4B and Qwen3.5-9B (Gated
    DeltaNet layers interleaved with periodic full-attention layers), $n{=}100$
    problems per benchmark. \emph{Single}: one-agent inference, shown for
    context. \emph{TextMAS}: agents exchange natural-language messages.
    \emph{\method{}}: agents exchange the KV cache of full-attention layers and
    the recurrent and causal-convolution states of Gated DeltaNet layers via the shared workspace. Metrics are accuracy (\%), output tokens per problem, and total inference time (s); lower is better for the latter
    two. \emph{Improve} is \method{} vs.\ TextMAS: accuracy in absolute points,
    token reduction (\%), speed-up (TextMAS time / \method{} time). Shaded
    \emph{Improve} cells mark a change in the favorable direction.}
  \label{tab:main-results}
  \centering
  \small   
  \setlength{\tabcolsep}{3pt}
  \begin{tabular}{ll cccc cccc}
    \toprule
    \multirow{2}{*}{\textbf{Tasks}} & \multirow{2}{*}{\textbf{Metrics}}
      & \multicolumn{4}{c}{\textbf{Qwen3.5-4B}}
      & \multicolumn{4}{c}{\textbf{Qwen3.5-9B}} \\
    \cmidrule(lr){3-6} \cmidrule(lr){7-10}
      & & Single & TextMAS & \method{} & Improve
        & Single & TextMAS & \method{} & Improve \\
    \midrule
    & Acc.   & 100.0 & 100.0 & 99.0 & \dn{1.0} & 100.0 & 98.0 & 99.0 & \gain{\up{1.0}} \\
    & Token  & 361 & 1138 & 361 & \gain{\dn{68.3\%}} & 386 & 1350 & 356 & \gain{\dn{73.6\%}} \\
    \multirow{-3}{*}{\textbf{ARC-E}} & Time (s) & 464 & 1294 & 846 & \gain{\spd{1.53}} & 426 & 1269 & 618 & \gain{\spd{2.05}} \\
    \midrule
    & Acc.   & 94.0 & 92.0 & 91.0 & \dn{1.0} & 94.0 & 95.0 & 93.0 & \dn{2.0} \\
    & Token  & 436 & 1297 & 481 & \gain{\dn{62.9\%}} & 465 & 1524 & 430 & \gain{\dn{71.8\%}} \\
    \multirow{-3}{*}{\textbf{ARC-C}} & Time (s) & 666 & 1881 & 1205 & \gain{\spd{1.56}} & 572 & 1588 & 739 & \gain{\spd{2.15}} \\
    \midrule
    & Acc.   & 92.0 & 85.0 & 93.0 & \gain{\up{8.0}} & 96.0 & 90.0 & 97.0 & \gain{\up{7.0}} \\
    & Token  & 393 & 898 & 286 & \gain{\dn{68.1\%}} & 368 & 1222 & 354 & \gain{\dn{71.0\%}} \\
    \multirow{-3}{*}{\textbf{GSM8K}} & Time (s) & 893 & 1361 & 846 & \gain{\spd{1.61}} & 476 & 1341 & 656 & \gain{\spd{2.04}} \\
    \midrule
    & Acc.   & 80.0 & 81.0 & 77.0 & \dn{4.0} & 89.0 & 88.0 & 88.0 & 0.0 \\
    & Token  & 1064 & 2139 & 907 & \gain{\dn{57.6\%}} & 1139 & 2740 & 966 & \gain{\dn{64.7\%}} \\
    \multirow{-3}{*}{\textbf{MedQA}} & Time (s) & 2061 & 3277 & 2815 & \gain{\spd{1.16}} & 1779 & 3878 & 2437 & \gain{\spd{1.59}} \\
    \midrule
    & Acc.   & 73.0 & 75.0 & 73.0 & \dn{2.0} & 85.0 & 79.0 & 76.0 & \dn{3.0} \\
    & Token  & 442 & 1219 & 369 & \gain{\dn{69.7\%}} & 723 & 1716 & 618 & \gain{\dn{64.0\%}} \\
    \multirow{-3}{*}{\textbf{MBPP+}} & Time (s) & 1633 & 2949 & 1625 & \gain{\spd{1.81}} & 1373 & 2770 & 1717 & \gain{\spd{1.61}} \\
    \midrule
    & Acc.   & 70.0 & 77.0 & 87.0 & \gain{\up{10.0}} & 83.0 & 93.0 & 90.0 & \dn{3.0} \\
    & Token  & 281 & 1283 & 563 & \gain{\dn{56.2\%}} & 617 & 1557 & 725 & \gain{\dn{53.4\%}} \\
    \multirow{-3}{*}{\textbf{HumanEval+}} & Time (s) & 802 & 2329 & 1606 & \gain{\spd{1.45}} & 1120 & 1894 & 1633 & \gain{\spd{1.16}} \\
    \bottomrule
  \end{tabular}
\end{table}

%

%% file: tables/small_model_results.tex
\providecommand{\method}{HyLaMAS}
\definecolor{winB}{HTML}{BFE3BF}
\providecommand{\gain}[1]{\cellcolor{winB}\textbf{#1}}

\begin{table}[t]
  \caption{Latent collaboration at small scale (Qwen3.5-2B), $n{=}100$ per
    benchmark. \emph{Single} is one-agent inference (context), \emph{TextMAS}
    passes natural-language messages, \method{} passes hybrid state. \emph{Improve}
    is \method{} vs.\ TextMAS (accuracy in points, tokens as \% reduction, speed as
    speed-up); shaded where it favors \method{}.}
  \label{tab:smallmodel-results}
  \centering
  \small
  \setlength{\tabcolsep}{4.5pt}
  \renewcommand{\arraystretch}{1.05}
  \begin{tabular}{l cccccc}
    \toprule
    \textbf{Method} & \textbf{GSM8K} & \textbf{ARC-E} & \textbf{ARC-C}
      & \textbf{MedQA} & \textbf{MBPP+} & \textbf{HumanEval+} \\
    \midrule
    \multicolumn{7}{l}{\textit{Accuracy (\%)} $\uparrow$} \\
    Single           & 73 & 96 & 87 & 61 & 42 & 50 \\
    TextMAS          & 67 & 91 & 81 & 61 & 49 & 52 \\
    \method{}        & 80 & 94 & 81 & 63 & 49 & 58 \\
    \textit{Improve} & \gain{+13} & \gain{+3} & +0 & \gain{+2} & +0 & \gain{+6} \\
    \midrule
    \multicolumn{7}{l}{\textit{Tokens / sample} $\downarrow$} \\
    Single           & 453 & 412 & 511 & 1137 & 570 & 608 \\
    TextMAS          & 1500 & 1249 & 1505 & 2852 & 1992 & 1869 \\
    \method{}        & 308 & 268 & 381 & 739 & 399 & 404 \\
    \textit{Improve} & \gain{$-$79.4\%} & \gain{$-$78.6\%} & \gain{$-$74.7\%}
                     & \gain{$-$74.1\%} & \gain{$-$80.0\%} & \gain{$-$78.4\%} \\
    \midrule
    \multicolumn{7}{l}{\textit{Total Inference time (s)} $\downarrow$} \\
    Single           & 718 & 423 & 889 & 2045 & 1797 & 1462 \\
    TextMAS          & 2305 & 1397 & 1853 & 4451 & 4126 & 3428 \\
    \method{}        & 830 & 514 & 681 & 2121 & 1745 & 1519 \\
    \textit{Improve} & \gain{2.78$\times$} & \gain{2.72$\times$} & \gain{2.72$\times$}
                     & \gain{2.10$\times$} & \gain{2.36$\times$} & \gain{2.26$\times$} \\
    \bottomrule
  \end{tabular}
\end{table}

%

%% file: appendix_revised.tex

\section{Appendix}
\label{sec:ablations}
\label{app:ablations}
\subsection{Ablations}
All ablations use deterministic decoding, seed 42, and the sequential
Planner--Critic--Refiner--Judger protocol. Only the Judger decodes text.
SelectKV is evaluated on Qwen3-4B with $n=100$, 10 latent steps, and 90\%
retention. The hybrid-channel ablation uses Qwen3.5-9B on MedQA with $n=100$,
batch size 8, 40 latent steps per non-Judger agent, and a 4,096-token generation
cap. Unlike the main experiments, this ablation disables latent-space
realignment. Exact commands are included in the supplementary repository.

Source code, experimental configurations, and reproduction commands
are available in the
\href{https://github.com/mliao2710/SelectKV}
{\selectkv{} repository} and the \href{https://github.com/aaron-sz/HyLaMAS}
{\hmethod{} repository}.

\begin{figure}[b]
  \centering
  \includegraphics[width=\linewidth]{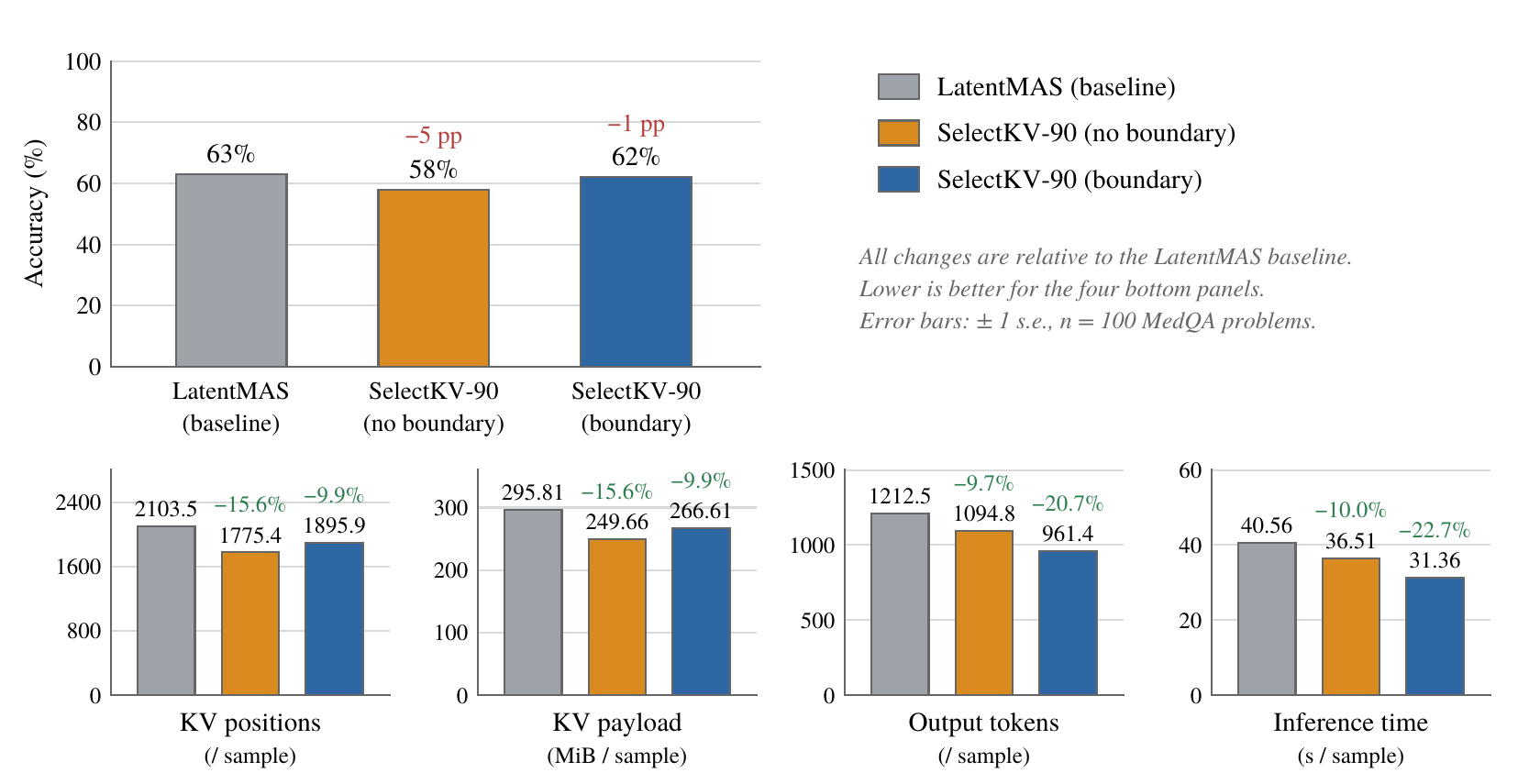}
  \caption{\textbf{Boundary-protection ablation for SelectKV on MedQA.}
  Removing protection permits greater compression but produces a larger
  observed accuracy loss. Boundary-aware SelectKV remains one accuracy point
  below full-cache LatentMAS while reducing communicated KV payload.}
  \label{fig:selectkv-ablation}
\end{figure}

We first evaluate the role of boundary protection in SelectKV. As shown in
Figure~\ref{fig:selectkv-ablation}, boundary-aware SelectKV improves MedQA
accuracy from 58\% to 62\% relative to the same 90\%-retention configuration
without boundary protection, compared with 63\% for full-cache LatentMAS.
At the same time, the boundary-aware configuration reduces KV payload from
295.81 MiB to 266.61 MiB per problem relative to the full-cache baseline.
These results suggest that preserving inherited context recovers much of the
accuracy lost under more aggressive compression while retaining a meaningful
communication reduction. 

\begin{figure}[t]
  \centering
  \includegraphics[width=\linewidth]{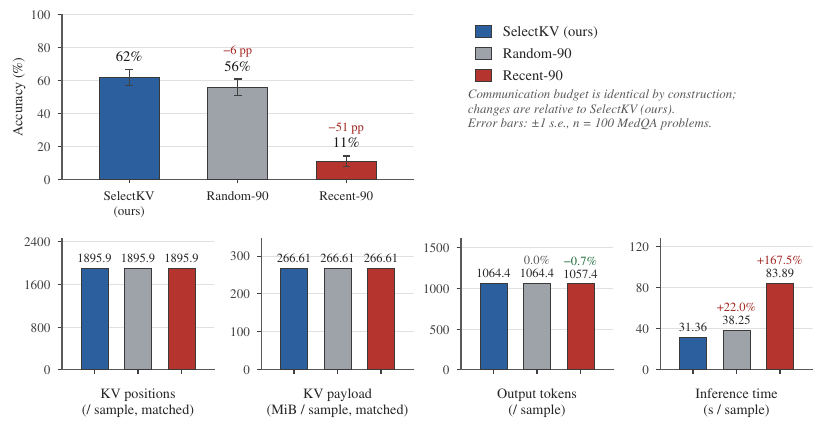}
  \caption{\textbf{Matched-budget SelectKV selector ablation on MedQA.}
  SelectKV outperforms random and recency-only selection at the same retained-KV
  budget, indicating that its selection criterion contributes beyond
  compression alone. The selector ablation was conducted as a separate experimental run
from the main SelectKV evaluation in Table~\ref{tab:selectkv-results}.}
  \label{fig:selectkv-selector-ablation}
\end{figure}

We next isolate the effect of the SelectKV ranking criterion from the
communication budget. Figure~\ref{fig:selectkv-selector-ablation} compares
SelectKV with random and recency-only selection under an identical retained-KV
budget. SelectKV obtains 62\% MedQA accuracy, compared with 56\% for random
selection and 11\% for recency-only selection. Because all three methods retain
the same number of KV positions, this ablation indicates that the SelectKV
selection criterion contributes beyond compression alone.

\begin{figure}[H]
  \centering
  \includegraphics[width=\linewidth]{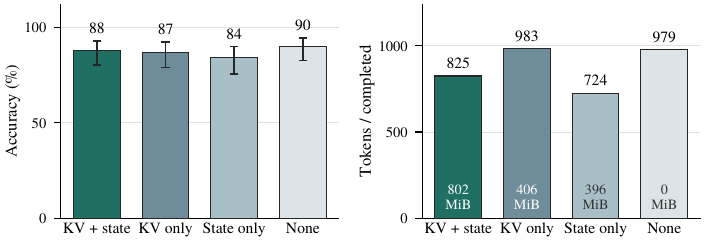}
  \caption{\textbf{Hybrid-state channel ablation for \hmethod{} on MedQA.}
  \emph{Both} retains full-attention KV and the complete GDN state, including
  recurrent and causal-convolution states; \emph{KV only} removes GDN state;
  \emph{State only} removes KV; and \emph{None} removes both. The left panel
  reports accuracy with approximate 95\% binomial confidence intervals. The
  right panel reports mean decoded tokens per completed problem and reports the logical state payload shared across agent boundaries in MiB.}
  \label{fig:channel-ablation}
\end{figure}

Finally, we separate the two communication channels used by \hmethod{}.
As shown in Figure~\ref{fig:channel-ablation}, accuracy is 88\% with both
full-attention KV and complete Gated DeltaNet (GDN) state, 87\% with KV only,
84\% with GDN state only, and 90\% with neither channel. None of the differences
from full-state transfer is significant under paired McNemar tests
($p=1.000$, $0.344$, and $0.754$, respectively). The right panel of
Figure~\ref{fig:channel-ablation} nevertheless shows a consistent descriptive
reduction in completed-response length when GDN state is retained: 825 versus
983 decoded tokens when KV is present, and 724 versus 979 when KV is removed.
Although the no-transfer configuration achieves 90\% accuracy versus 88\% with both, full-state transfer reduces decoded tokens from 979 to 825 per completed response (15.7\% fewer), suggesting an output-length benefit without a demonstrated accuracy advantage.


\providecommand{\TODO}[1]{\textcolor{red}{[#1]}}

\subsection{Experimental Setup and Reproducibility}
\label{app:setup}

\paragraph{Protocol.}
Both methods use the sequential Planner--Critic--Refiner--Judger chain. The
first three agents append a role-conditioned prompt and perform latent rollout
steps without decoding text; the Judger receives the accumulated state, appends
its prompt, and decodes the answer. All agents in a run share one model
instance, a single set of weights, no adapters, no per-role fine-tuning, 
so they differ only in their prompt and in the state they inherit. Padding-aware
position and attention masks keep batched examples independent.

\begin{table}[H]
  \centering
  \small
  \setlength{\tabcolsep}{5pt}
  \caption{Experimental configurations for \selectkv{} and \hmethod{}. Runtime comparisons use matched model, benchmark, hardware, and batch-size settings.}
  \label{tab:setup}
  \begin{tabular}{@{}lll@{}}
    \toprule
    & \textbf{\selectkv{}} & \textbf{\hmethod{}} \\
    \midrule
    Model            & Qwen3-4B & Qwen3.5-\{2B, 4B, 9B\} \\
    Architecture     & dense attention, 36 layers & GDN\,:\,attention $3{:}1$ \\
    Weights          & \texttt{bfloat16} & \texttt{bfloat16} \\
    Attention backend & --- & \texttt{sdpa} \\
    GPU              & NVIDIA L40S (46,068 MiB) & A100-SXM4-40GB (2B/4B); L40S (9B) \\
    PyTorch / CUDA   & 2.13.0 / 13.0; 2.8.0 / 12.9 & 2.13.0 / 13.0 (2B/4B); 2.8.0 / 12.9 (9B) \\
    \texttt{transformers} & 5.15.1 & 5.15.1 \\
    Latent steps $m$ & 10 & 40 \\
    Batch size       & 1 & 8 \\
    Retention $r$    & 0.90 & --- \\
    Decoding         & \multicolumn{2}{l}{\texttt{do\_sample=False}, seed 42} \\
    Examples         & \multicolumn{2}{l}{$n=100$ per benchmark, offset 0} \\
    \bottomrule
  \end{tabular}
\end{table}

\paragraph{Code and Reproducibility.}
Source code, experimental configurations, and reproduction commands
are available in the
\href{https://github.com/mliao2710/SelectKV}
{\selectkv{} repository}
and the
\href{https://github.com/aaron-sz/HyLaMAS}
{\hmethod{} repository}.
Table~\ref{tab:setup} summarizes the key experimental settings. The repositories provide detailed implementation and reproduction
instructions.

%% file: sources.bib
@misc{zou2026latentcollaborationmultiagentsystems,
      title={Latent Collaboration in Multi-Agent Systems}, 
      author={Jiaru Zou and Ruizhong Qiu and Gaotang Li and Xiyuan Yang and Katherine Tieu and Pan Lu and Ke Shen and Hanghang Tong and Yejin Choi and Jingrui He and James Zou and Mengdi Wang and Ling Yang},
      year={2026},
      eprint={2511.20639},
      archivePrefix={arXiv},
      primaryClass={cs.CL},
      url={https://arxiv.org/abs/2511.20639}, 
}

@misc{li2026latentleadsbetterrelay,
      title={When Less Latent Leads to Better Relay: Information-Preserving Compression for Latent Multi-Agent LLM Collaboration}, 
      author={Yiping Li and Zhiyu An and Wan Du},
      year={2026},
      eprint={2604.13349},
      archivePrefix={arXiv},
      primaryClass={cs.LG},
      url={https://arxiv.org/abs/2604.13349}, 
}

@misc{shi2026kvcommenablingefficientllm,
      title={KVComm: Enabling Efficient LLM Communication through Selective KV Sharing}, 
      author={Xiangyu Shi and Marco Chiesa and Gerald Q. Maguire Jr. and Dejan Kostic},
      year={2026},
      eprint={2510.03346},
      archivePrefix={arXiv},
      primaryClass={cs.LG},
      url={https://arxiv.org/abs/2510.03346}, 
}

@misc{ye2025kvcommonlinecrosscontextkvcache,
      title={KVCOMM: Online Cross-context KV-cache Communication for Efficient LLM-based Multi-agent Systems}, 
      author={Hancheng Ye and Zhengqi Gao and Mingyuan Ma and Qinsi Wang and Yuzhe Fu and Ming-Yu Chung and Yueqian Lin and Zhijian Liu and Jianyi Zhang and Danyang Zhuo and Yiran Chen},
      year={2025},
      eprint={2510.12872},
      archivePrefix={arXiv},
      primaryClass={cs.MA},
      url={https://arxiv.org/abs/2510.12872}, 
}

@misc{liu2023scissorhandsexploitingpersistenceimportance,
      title={Scissorhands: Exploiting the Persistence of Importance Hypothesis for LLM KV Cache Compression at Test Time}, 
      author={Zichang Liu and Aditya Desai and Fangshuo Liao and Weitao Wang and Victor Xie and Zhaozhuo Xu and Anastasios Kyrillidis and Anshumali Shrivastava},
      year={2023},
      eprint={2305.17118},
      archivePrefix={arXiv},
      primaryClass={cs.LG},
      url={https://arxiv.org/abs/2305.17118}, 
}

@misc{zhang2023h2oheavyhitteroracleefficient,
      title={H$_2$O: Heavy-Hitter Oracle for Efficient Generative Inference of Large Language Models}, 
      author={Zhenyu Zhang and Ying Sheng and Tianyi Zhou and Tianlong Chen and Lianmin Zheng and Ruisi Cai and Zhao Song and Yuandong Tian and Christopher Ré and Clark Barrett and Zhangyang Wang and Beidi Chen},
      year={2023},
      eprint={2306.14048},
      archivePrefix={arXiv},
      primaryClass={cs.LG},
      url={https://arxiv.org/abs/2306.14048}, 
}

@misc{xiao2024efficientstreaminglanguagemodels,
      title={Efficient Streaming Language Models with Attention Sinks}, 
      author={Guangxuan Xiao and Yuandong Tian and Beidi Chen and Song Han and Mike Lewis},
      year={2024},
      eprint={2309.17453},
      archivePrefix={arXiv},
      primaryClass={cs.CL},
      url={https://arxiv.org/abs/2309.17453}, 
}

@misc{cobbe2021trainingverifierssolvemath,
      title={Training Verifiers to Solve Math Word Problems}, 
      author={Karl Cobbe and Vineet Kosaraju and Mohammad Bavarian and Mark Chen and Heewoo Jun and Lukasz Kaiser and Matthias Plappert and Jerry Tworek and Jacob Hilton and Reiichiro Nakano and Christopher Hesse and John Schulman},
      year={2021},
      eprint={2110.14168},
      archivePrefix={arXiv},
      primaryClass={cs.LG},
      url={https://arxiv.org/abs/2110.14168}, 
}

@misc{knnlatentmas,
      title={kNN-latentMAS: LatentMAS with kNN KV Cache Pruning}, 
      author={Lawrence Liu},
      year={2026},
      url={https://github.com/Bookmaster9/kNN-latentMAS}, 
}

@misc{jin2020diseasedoespatienthave,
      title={What Disease does this Patient Have? A Large-scale Open Domain Question Answering Dataset from Medical Exams}, 
      author={Di Jin and Eileen Pan and Nassim Oufattole and Wei-Hung Weng and Hanyi Fang and Peter Szolovits},
      year={2020},
      eprint={2009.13081},
      archivePrefix={arXiv},
      primaryClass={cs.CL},
      url={https://arxiv.org/abs/2009.13081}, 
}

@misc{clark2018thinksolvedquestionanswering,
      title={Think you have Solved Question Answering? Try ARC, the AI2 Reasoning Challenge}, 
      author={Peter Clark and Isaac Cowhey and Oren Etzioni and Tushar Khot and Ashish Sabharwal and Carissa Schoenick and Oyvind Tafjord},
      year={2018},
      eprint={1803.05457},
      archivePrefix={arXiv},
      primaryClass={cs.AI},
      url={https://arxiv.org/abs/1803.05457}, 
}

@misc{liu2023codegeneratedchatgptreally,
      title={Is Your Code Generated by ChatGPT Really Correct? Rigorous Evaluation of Large Language Models for Code Generation}, 
      author={Jiawei Liu and Chunqiu Steven Xia and Yuyao Wang and Lingming Zhang},
      year={2023},
      eprint={2305.01210},
      archivePrefix={arXiv},
      primaryClass={cs.SE},
      url={https://arxiv.org/abs/2305.01210}, 
}

@misc{yang2025gateddeltanetworksimproving,
      title={Gated Delta Networks: Improving Mamba2 with Delta Rule}, 
      author={Songlin Yang and Jan Kautz and Ali Hatamizadeh},
      year={2025},
      eprint={2412.06464},
      archivePrefix={arXiv},
      primaryClass={cs.CL},
      url={https://arxiv.org/abs/2412.06464}, 
}

@misc{qwen35blog,
    title = {Qwen3.5: Towards Native Multimodal Agents},
    url = {https://qwen.ai/blog?id=qwen3.5},
    author = {Qwen Team},
    month = {February},
    year = {2026}
}

@misc{zhang2024chainagentslargelanguage,
      title={Chain of Agents: Large Language Models Collaborating on Long-Context Tasks}, 
      author={Yusen Zhang and Ruoxi Sun and Yanfei Chen and Tomas Pfister and Rui Zhang and Sercan Ö. Arik},
      year={2024},
      eprint={2406.02818},
      archivePrefix={arXiv},
      primaryClass={cs.CL},
      url={https://arxiv.org/abs/2406.02818}, 
}

@misc{wu2023autogenenablingnextgenllm,
      title={AutoGen: Enabling Next-Gen LLM Applications via Multi-Agent Conversation}, 
      author={Qingyun Wu and Gagan Bansal and Jieyu Zhang and Yiran Wu and Beibin Li and Erkang Zhu and Li Jiang and Xiaoyun Zhang and Shaokun Zhang and Jiale Liu and Ahmed Hassan Awadallah and Ryen W White and Doug Burger and Chi Wang},
      year={2023},
      eprint={2308.08155},
      archivePrefix={arXiv},
      primaryClass={cs.AI},
      url={https://arxiv.org/abs/2308.08155}, 
}

@misc{du2023improvingfactualityreasoninglanguage,
      title={Improving Factuality and Reasoning in Language Models through Multiagent Debate}, 
      author={Yilun Du and Shuang Li and Antonio Torralba and Joshua B. Tenenbaum and Igor Mordatch},
      year={2023},
      eprint={2305.14325},
      archivePrefix={arXiv},
      primaryClass={cs.CL},
      url={https://arxiv.org/abs/2305.14325}, 
}

@misc{liang2024encouragingdivergentthinkinglarge,
      title={Encouraging Divergent Thinking in Large Language Models through Multi-Agent Debate}, 
      author={Tian Liang and Zhiwei He and Wenxiang Jiao and Xing Wang and Yan Wang and Rui Wang and Yujiu Yang and Shuming Shi and Zhaopeng Tu},
      year={2024},
      eprint={2305.19118},
      archivePrefix={arXiv},
      primaryClass={cs.CL},
      url={https://arxiv.org/abs/2305.19118}, 
}

@misc{pham2024letmodelsspeakciphers,
      title={Let Models Speak Ciphers: Multiagent Debate through Embeddings}, 
      author={Chau Pham and Boyi Liu and Yingxiang Yang and Zhengyu Chen and Tianyi Liu and Jianbo Yuan and Bryan A. Plummer and Zhaoran Wang and Hongxia Yang},
      year={2024},
      eprint={2310.06272},
      archivePrefix={arXiv},
      primaryClass={cs.CL},
      url={https://arxiv.org/abs/2310.06272}, 
}

@misc{fu2026cachetocachedirectsemanticcommunication,
      title={Cache-to-Cache: Direct Semantic Communication Between Large Language Models}, 
      author={Tianyu Fu and Zihan Min and Hanling Zhang and Jichao Yan and Guohao Dai and Wanli Ouyang and Yu Wang},
      year={2026},
      eprint={2510.03215},
      archivePrefix={arXiv},
      primaryClass={cs.CL},
      url={https://arxiv.org/abs/2510.03215}, 
}

@misc{ramesh2025communicatingactivationslanguagemodel,
      title={Communicating Activations Between Language Model Agents}, 
      author={Vignav Ramesh and Kenneth Li},
      year={2025},
      eprint={2501.14082},
      archivePrefix={arXiv},
      primaryClass={cs.CL},
      url={https://arxiv.org/abs/2501.14082}, 
}

@misc{gu2024mambalineartimesequencemodeling,
      title={Mamba: Linear-Time Sequence Modeling with Selective State Spaces}, 
      author={Albert Gu and Tri Dao},
      year={2024},
      eprint={2312.00752},
      archivePrefix={arXiv},
      primaryClass={cs.LG},
      url={https://arxiv.org/abs/2312.00752}, 
}

@misc{dao2024transformersssmsgeneralizedmodels,
      title={Transformers are SSMs: Generalized Models and Efficient Algorithms Through Structured State Space Duality}, 
      author={Tri Dao and Albert Gu},
      year={2024},
      eprint={2405.21060},
      archivePrefix={arXiv},
      primaryClass={cs.LG},
      url={https://arxiv.org/abs/2405.21060}, 
}

@misc{yang2024gatedlinearattentiontransformers,
      title={Gated Linear Attention Transformers with Hardware-Efficient Training}, 
      author={Songlin Yang and Bailin Wang and Yikang Shen and Rameswar Panda and Yoon Kim},
      year={2024},
      eprint={2312.06635},
      archivePrefix={arXiv},
      primaryClass={cs.LG},
      url={https://arxiv.org/abs/2312.06635}, 
}

@misc{zuo2024falconmambacompetitiveattentionfree,
      title={Falcon Mamba: The First Competitive Attention-free 7B Language Model}, 
      author={Jingwei Zuo and Maksim Velikanov and Dhia Eddine Rhaiem and Ilyas Chahed and Younes Belkada and Guillaume Kunsch and Hakim Hacid},
      year={2024},
      eprint={2410.05355},
      archivePrefix={arXiv},
      primaryClass={cs.CL},
      url={https://arxiv.org/abs/2410.05355}, 
}

@misc{lieber2024jambahybridtransformermambalanguage,
      title={Jamba: A Hybrid Transformer-Mamba Language Model}, 
      author={Opher Lieber and Barak Lenz and Hofit Bata and Gal Cohen and Jhonathan Osin and Itay Dalmedigos and Erez Safahi and Shaked Meirom and Yonatan Belinkov and Shai Shalev-Shwartz and Omri Abend and Raz Alon and Tomer Asida and Amir Bergman and Roman Glozman and Michael Gokhman and Avashalom Manevich and Nir Ratner and Noam Rozen and Erez Shwartz and Mor Zusman and Yoav Shoham},
      year={2024},
      eprint={2403.19887},
      archivePrefix={arXiv},
      primaryClass={cs.CL},
      url={https://arxiv.org/abs/2403.19887}, 
}
